# EURADOS Intercomparison on the Usage of the ICRP/ICRU Adult Reference Computational Phantoms

Maria Zankl[1,*], Jonathan Eakins[2], José-María Gómez Ros[3], Christelle Huet[4], Jan Jansen[2], Montserrat Moraleda[3], Uwe Reichelt[5], Lara Struelens[6], Tomas Vrba[7]

[1] Helmholtz Zentrum München (GmbH) German Research Center for Environmental Health, Institute of Radiation Medicine, Ingolstädter Landstr. 1, 85764 Neuherberg, Germany
[2] Public Health England (PHE), Chilton, UK
[3] CIEMAT – Centro de Investigaciones Energéticas, Medioambientales y Tecnológicas, Madrid, Spain
[4] Institut de Radioprotection et de Sûreté Nucléaire, Fontenay-aux-Roses, France
[5] Technical University Dresden, Dresden, Germany
[6] SCK CEN – Belgian Nuclear Research Centre, Mol, Belgium
[7] Faculty of Nuclear Sciences and Physical Engineering, Czech Technical University in Prague, Prague, Czech Republic
*Corresponding author: zankl@helmholtz-muenchen.de

*Abstract*—The European Radiation Dosimetry Group, EURADOS, has organised an intercomparison study on the usage of the ICRP/ICRU voxel reference computational phantoms together with radiation transport codes. Voluntary participants have been invited to solve specific tasks and provide solutions to the organisers before a certain deadline. The tasks to be solved are of practical interest in occupational, environmental and medical dosimetry. The aims of this training activity were to investigate if the phantoms have been correctly implemented in the radiation transport codes and to give the participants the opportunity to check their own calculations against quality-assured master solutions and improve their approach, if needed.

Keywords — intercomparison; reference computational phantoms; radiation transport codes; training

1. Introduction

EURADOS, the European Radiation Dosimetry Group, is a network of more than 75 European institutions and 600 scientists coordinated in working groups that – among other activities – organises scientific research meetings and training activities as well as intercomparison and benchmark studies.

Since most radiation transport codes are rather complex, many – especially novice – users are applying them as "black boxes", sometimes failing to realise whether the parameters chosen are indeed suitable for the tasks to be solved. This is one of the reasons why EURADOS aims at improving this situation by organising intercomparison studies (Broggio et al., 2012; Gómez-Ros et al., 2008; Gualdrini et al., 2005; Price et al., 2006; Siebert et al., 2006; Tanner et al., 2004; Vrba et al., 2015; Vrba et al., 2014), in which participants are invited to solve proposed computational tasks and check their results against both quality-assured so-called "master solutions" provided by EURADOS and the solutions of other participants.

EURADOS Working Group 6 "Computational Dosimetry" recently organised an intercomparison study on the usage of the ICRP/ICRU adult reference computational phantoms (ICRP, 2009) that aimed to investigate whether participants were able to correctly combine the phantoms with the radiation transport codes used, and if they were able to correctly apply ICRP guidance on the evaluation of specific dosimetric quantities such as organ absorbed and/or equivalent dose (in particular for the red bone marrow) (ICRP, 2010) and effective dose (ICRP, 2007). The purpose of this article is to summarise the general aspects of the intercomparison exercise.

2. Methods and Materials

2.1 Phantoms

Two phantoms of the human body were to be used in the intercomparison exercise. These are the male and female adult reference computational phantoms as described in ICRP Publication 110 (ICRP, 2009). The phantoms are based on the voxel models "Golem" (Zankl and Wittmann, 2001) and "Laura" (Zankl et al., 2005), which are in turn based on medical image data of real people whose body height and mass resembled the reference anatomical and physiological parameters for both male and female subjects given in Publication 89 (ICRP, 2002). For construction of the reference phantoms, several modification steps were applied to the segmented phantoms Golem and Laura. These were:

- voxel scaling to match reference height and reference skeleton mass;
- inclusion of further anatomical details, such as a greater amount of blood vessels, bronchi, and lymphatic nodes;
- sub-segmentation of the skeleton;
- matching the organ masses of both models to the ICRP data on the adult Reference Male and Reference Female without compromising their anatomic realism;
- adjusting the whole-body masses to 73 and 60 kg for the male and female reference computational phantoms, respectively, by "wrapping" the body with additional layers of adipose tissue.

The adult male and female reference computational phantoms are shown in Figure 1, and their main characteristics are summarised in Table 1.

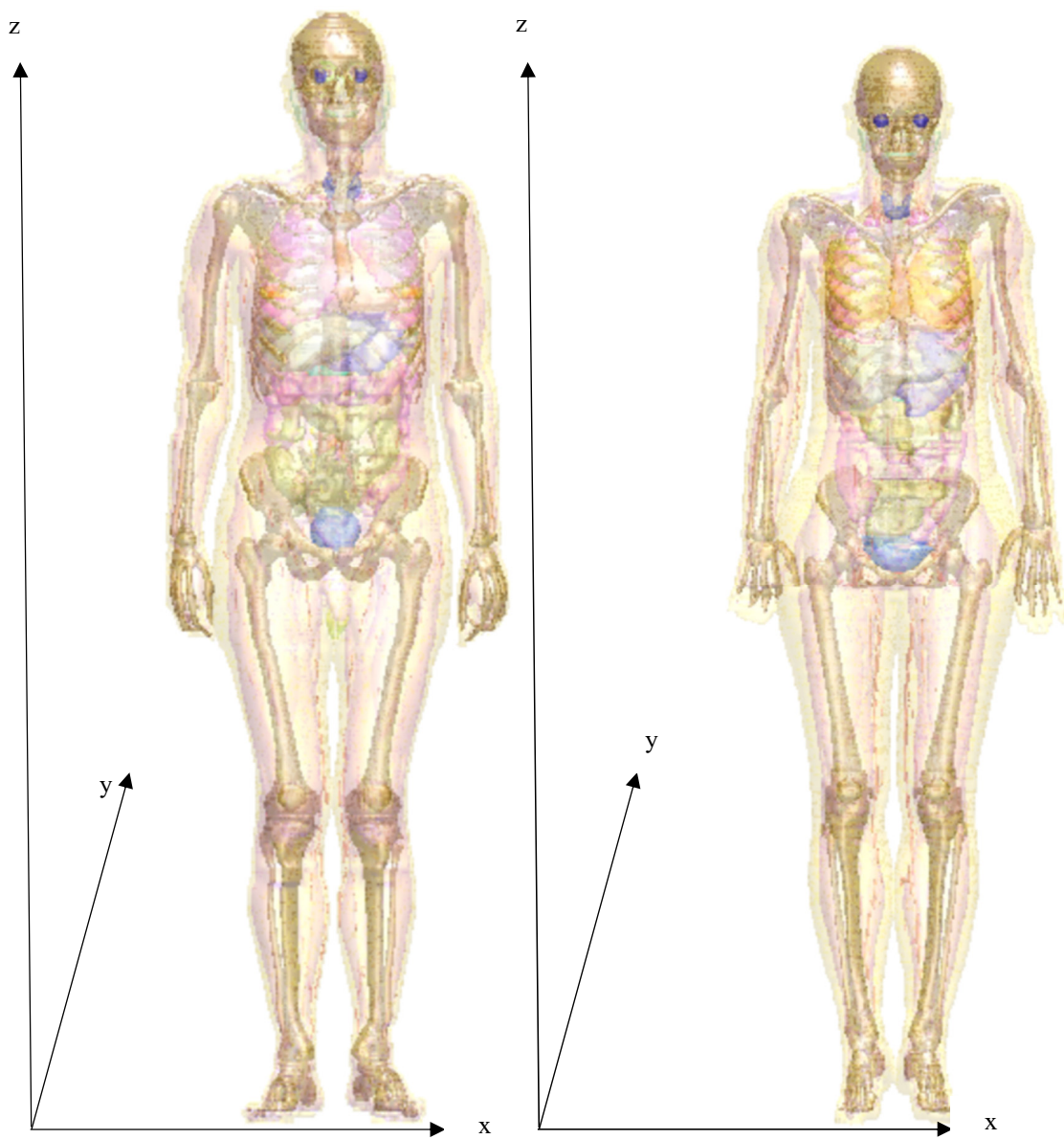

**Figure 1: Adult male (left) and female (right) computational reference phantoms together with their coordinate system. The x axis is represented by the columns of the voxel array and is directed from the phantoms' right to their left side, the y axis is represented by the rows and is directed from the phantoms' front to their back, and the z axis is represented by the array slices and is directed from the feet towards the head.**

**Table 1:** Main characteristics of the adult male and female reference computational phantoms (ICRP, 2009)

| Property | Male | Female |
|---|---|---|
| Height (m) | 1.76 | 1.63 |
| Mass (kg) | 73.0 | 60.0 |
| Number of tissue voxels | 1946375 | 3886020 |

| Slice thickness (voxel height, mm) | 8.0 | 4.84 |
|---|---|---|
| Voxel in-plane resolution (mm) | 2.137 | 1.775 |
| Voxel volume (mm$^3$) | 36.54 | 15.25 |
| Number of columns | 254 | 299 |
| Number of rows | 127 | 137 |
| Number of slices | 220 (+2)$^{a)}$ | 346 (+2)$^{a)}$ |

a) Additional slices of skin at the top and bottom to cover structures other than skin of the head and feet in the top and bottom slices

The human skeleton is composed of cortical bone, trabecular bone, active (red) and inactive (yellow) bone marrow, cartilage and endosteal tissues. For purposes of radiological protection, the ICRP defines two skeletal cell populations of dosimetric interest relevant to stochastic biological effects: (1) haematopoietic stem cells associated with the risk of radiogenic leukaemia, and (2) osteoprogenitor cells associated with the risk of radiogenic bone cancer. The former are represented by the active bone marrow, the latter by the total marrow (active and inactive) within 50 μm distance from the bone surfaces. The dimensions of internal structures of the skeletal tissues are in the order of micrometres and therefore much smaller than the resolution of a standard medical computed tomography (CT) scan which is typically around millimeters. Thus, these volumes could not be 'segmented', i.e. resolved into distinct voxelized regions. To account for this circumstance, a specific method for bone dosimetry has been developed and introduced in ICRP Publication 116 (ICRP, 2010), based on bone- and energy-specific fluence-to-dose response functions for photons and neutrons, or the use of equations to obtain the skeletal-averaged absorbed dose to active marrow and endosteum in the case of directly ionizing radiation, such as electrons.

The participants were advised to use the bone dosimetry method as suggested by ICRP Publication 116 (ICRP, 2010). Furthermore, they were asked to use the reference computational phantoms with their organ masses and tissue compositions as described in ICRP Publication 110 (ICRP, 2009), in contrast to the organ masses with added blood content as introduced in ICRP Publication 133 (ICRP, 2016).

The phantom data are given as an ASCII file consisting of an array of organ identification numbers listed slice by slice; within each slice, row by row; and within each row, column by column (ICRP, 2009). The elemental composition and density for the material assigned to every organ ID are also provided. The data are publically available as online supplement of ICRP Publication 110.

## 2.2 Skills to be tested

Participants were invited to attempt to solve the tasks and submit their results to the organisers. Besides testing several skills of the participants, further aims of the intercomparison exercise were:
- to provide an opportunity for the participants to improve their computational procedures via feedback,
- to identify common pitfalls, and
- to gain insight into the status of voxel phantom usage in computational dosimetry.

Besides testing the correct usage of the Monte Carlo codes involved, the intercomparison exercise also aimed to find out if the participants
- were able to correctly combine the ICRP/ICRU reference computational phantoms of ICRP Publication 110 (ICRP, 2009) with their radiation transport codes,
- understood a variety of dose quantities, such as
  - organ absorbed dose,
  - organ absorbed dose rate,
  - organ equivalent dose,
  - organ equivalent dose rate,
  - absorbed fraction,
  - specific absorbed fraction,
  - S value,
  - and effective dose (ICRP, 2007),
- understood the correct application of a variety of normalisation quantities (e.g., air kerma free in air, kerma-area product, activity concentration),
- and were able to correctly apply the methods for red bone marrow and endosteum dosimetry as recommended in ICRP Publication 116 (ICRP, 2010).

## 2.3 Tasks to be solved

The tasks to be solved considered a variety of exposure scenarios (occupational, environmental and medical) and radiation types (photons, electrons, neutrons). More specifically, these were:
- A photon point source in front of the phantoms at 125 cm from the bottom of the feet and 100 cm from the chest: The aim was to calculate organ absorbed doses for both reference computational phantoms and the effective dose from a Co-60 source with activity of 1 GBq during ten minutes exposure time.

- A neutron point source in front of the phantoms at 125 cm from the bottom of the feet and 100 cm from the chest: The aim was to calculate organ absorbed doses for both reference computational phantoms and the effective dose for a 1 minute exposure to a 1 GBq source of 10 keV neutrons.

- Am-241 ground contamination: The contamination is assumed to be contained within a disc of radius 2 m, with the anthropomorphic phantom standing at its center, and is deposited on the surface of a concrete floor; a uniform ground contamination is assumed, with an emission rate of 1 photon per $cm^2$ per second. The aim was to calculate organ absorbed dose rates for both reference computational phantoms, as well as the effective dose rate.

- Immersion in a radionuclide source homogeneously distributed inside a room: The ICRP/ICRU adult male reference phantom is located at the center of a confined room filled with N-16 contaminated air. The aim was to calculate organ equivalent dose rates per activity concentration.

- Typical x-ray examinations: The aim of this exercise was to calculate organ absorbed dose conversion coefficients per air kerma and per kerma-area product for the male and female reference computational phantom for two typical x-ray examinations (chest PA and abdomen AP).

- Internal dosimetry: The aims of this exercise were to evaluate (1) absorbed fractions and specific absorbed fractions of energy in specified "target" organs for (1a) monoenergetic photons and (1b) monoenergetic electrons distributed homogeneously in specific "source" organs of both phantoms, and (2) S-values for the same source and target organ combinations for specific radionuclides.

## 2.4    Approach chosen

Each of the tasks was supervised by two or three members of EURADOS WG6. One person was responsible for providing a master solution, the correctness of which had to be ascertained by second/third calculations by the other members supporting the task.

A collection of the task specifications was announced on the EURADOS website (http://www.eurados.org/) and distributed to various mailing lists for recruiting potential participants in May 2018. Each interested participant was free to solve one or several of the problems, according to his/her knowledge, interest, and time to be devoted to the participation.

The participants had to provide their solutions to the person responsible for each specific task by a specified deadline. Microsoft Excel templates for entering the participants' solutions in a pre-defined format were provided in order to ease evaluation by the responsible persons. The templates contained also a general part asking for personal and affiliation details, as well as information about the transport code used and its version, the cross-section libraries, cutoff values chosen, the potential use of kerma approximation, and the method of bone dosimetry applied. In case the latter deviated from the ICRP 116 method (ICRP, 2010), participants were requested to explain their method in detail. The solutions were evaluated, and feedback to the participants was provided in spring and summer 2019; potential mistakes were aimed at being resolved by direct contact between the responsible persons and participants. The final deadline for revised solutions was at the end of May 2020.

Although the results are presented anonymously, all participants were invited to co-author the manuscripts that contain the detailed analyses of the results of the tasks to which they contributed, and some have accepted the offer. The respective articles are part of the present Special Issue of Radiation Measurements.

3. Response and general findings

The Intercomparison Exercise was well-received by the computational dosimetry community. 32 participants from 17 countries submitted solutions to at least one of the proposed tasks; some participants solved several or even all tasks. The agreement of the submitted solutions with the master solutions was very variable – ranging from excellent agreement to discrepancies of several orders of magnitude in single cases. Several participants were found to have had problems in correctly applying the ICRP recommended method of red bone marrow dosimetry using dose response functions (ICRP, 2010). This is the reason why there is a specific article in this Special Issue that describes this method in more detail (Zankl et al., 2021).

Many problems in the initial submissions could be solved by feedback between the participants and the persons responsible for each task. In most cases, the participants then resubmitted a revised set of results. Some initial errors were attributable to simple carelessness, such as copy-and-paste errors or mis-arranging the results in the given template. Sometimes there was a misunderstanding concerning the normalisation quantity, e.g., normalising to the correct quantity but at a different distance from the source than was asked for. These errors were mostly easy to find. There were, however, cases where the participants did not disclose how they changed their computational procedure to obtain a revised solution; in these cases, no knowledge about the nature of the initial mis-comprehensions can be gained, unfortunately.

One general finding was that some participants provided results that were obviously wrong, although this could have been revealed by simple plausibility checks. Such tests might have been performed directly on the results by checking whether the individual organ dose coefficients showed a reasonable pattern – e.g., for homogeneous irradiation conditions, all of the resulting organ dose coefficients should have a rather similar magnitude. Furthermore, for specific tasks, an indication of the reasonableness of the results might have been obtained by comparing the results with available literature values for exposure conditions that are not too different from the task considered. Such simple measures of quality assurance were sometimes neglected, however.

4. Conclusion

The tasks that were set in the EURADOS intercomparison exercise are of practical interest in the fields of medical physics as well as occupational and environmental radiation protection. A correct simulation of the proposed tasks with computer codes requires an appropriate knowledge of the physical quantities involved and the ability to combine the ICRP/ICRU reference computational phantoms correctly with radiation transport codes.

The main scope of the intercomparison exercise was to offer an open forum for discussion and training in the field of computational dosimetry. In many cases, initial errors made by the participants were easy to find and eliminate. In some other cases, however, no knowledge about potential miscomprehensions could be gained due to the participants not disclosing how they improved their computational procedure.

One general conclusion is also that there was sometimes a lack of awareness of the necessity to quality assure computational results, such as with the help of plausibility checks or comparison with literature data for similar exposure conditions.

The present intercomparison exercise demonstrated once more that these types of study are beneficial to the field of computational dosimetry. Besides training the participants directly by improving their computational procedures via feedback with the task organisers, they lead also to the availability of representative dose values for various exposure conditions that may aid future novice users in the quality assurance of their methods.

Acronyms

ICRP: International Commission on Radiological Protection
ICRU: International Commission on Radiation Units and Measurements